\documentclass{iopconfser}
\usepackage{amssymb, amsmath, amsthm}
\newtheorem{Theorem}{Theorem}[section]
\theoremstyle{plain} % or definition, remark, etc.
\usepackage{graphicx}
\usepackage{hyperref}
\newcommand{\mono}{{\mathbf{M}}}

\usepackage{orcidlink}
\begin{document}

\title{Hyperbolic Mass in 2+1 Dimensions}

\author{Raphaela Wutte$^{1}$\orcidlink{0000-0002-1346-1047}}

\affil{$^1$Department of Physics and Beyond: Center for Fundamental Concepts in Science, Arizona
State University}

\email{rwutte@hep.itp.tuwien.ac.at}

\begin{abstract}
This short review surveys mass for two-dimensional asymptotically locally hyperbolic initial data sets. I explain the difficulties in defining mass in spatial dimension two, which are resolved via minimisation using a positive energy theorem, and review how gluing theorems can be used to construct novel initial data sets with controlled mass.
\\ \\
Based on joint work with P.\ T.\ Chruściel

\end{abstract}

\section{Introduction}
Solutions to Einstein's equations with a negative cosmological constant are ubiquitous in modern physics. They have received a surge of attention due to the Anti-de
Sitter/conformal field theory correspondence — a conjecture which asserts that quantum
gravity on spacetimes which asymptote to Anti-de Sitter in $d$ spacetime dimensions, so-called \emph{asymptotically Anti-de Sitter} spacetimes, is equivalent to a quantum field
theory with conformal invariance in $d-1$ dimensions (AdS$_d$/CFT$_{d-1}$ correspondence). While the prototypical example of this conjecture is the well-studied AdS$_5$/CFT$_4$ correspondence \cite{Maldacena:1997re}, the three-dimensional AdS$_3$/CFT$_2$ correspondence has also garnered significant attention.

While much work in the physics community has been devoted to (2+1)-dimensional general relativity in recent decades, the existing literature on the general relativistic constraint equations has usually restricted itself to spatial dimensions $n = 3$ or
higher (see, e.g. \cite{BartnikIsenberg,CarlottoLRR} and references therein). This is presumably because of the fact that all $(2+1)$-dimensional vacuum spacetimes are locally isometric at fixed cosmological constant, suggesting that this might not be a worthwhile topic to study from a purely classical perspective.

However, global properties still allow for interesting solutions, such as the so-called \emph{BTZ black hole} \cite{Banados:1992wn}, a vacuum solution which has an event horizon despite being locally isometric to AdS$_3$. Moreover, as we will see here, the richness of the asymptotic structure \cite{Brown:1986nw}, leads to the mass of two-dimensional initial data sets having properties which are unlike the ones in higher dimensions \cite{Chrusciel:2024pvn}.

This proceeding is based on a talk given at GR24 \&
Amaldi16 about joint work with P.\ T.\ Chruściel \cite{Chrusciel:2024pvn}. I will also briefly comment on \cite{Chrusciel:2024vle}, joint work with P.\ T.\ Chruściel, W.\ Cong and T.\ Queau.

\section{Mass}
In spatial dimension two, vacuum and time symmetry (vanishing extrinsic curvature), the general relativistic constraint equations read
\begin{equation}
\label{constrainteq}
    R = - 2\,,
\end{equation}
where we have normalised the cosmological constant $\Lambda$ to $-1$. In two dimensions, hyperbolic space is, up to isometry, the unique, simply-connected manifold with constant scalar curvature $R = -2$.
Here, we will be interested in metrics which asymptote to hyperbolic space at large distances, which we will call \emph{asymptotically locally hyperbolic}. 

We start by considering an asymptotically locally hyperbolic metric in the Fefferman-Graham coordinate system 
\begin{equation}
  g = r^{-2} dr^2 +
  r^2  g_{\varphi\varphi} (r, \varphi)  d\varphi^2
  \,.
 \end{equation}
Expanding $g_{\varphi\varphi}(r, \varphi)$ in powers of $r$ at large $r$
 and solving the constraint equations (\ref{constrainteq}),
 one finds that the expansion stops
\begin{equation}
  g = r^{-2}  dr^2 + \big(
   r^2 + \frac{\mu(\varphi)}{2}   +  \frac{\mu(\varphi)^2}{16 r^2}
   \big)
    d\varphi^2\,,
    \label{banados}
 \end{equation}
 a fact which has been already pointed out in \cite{Skenderis:1999nb, Banados:1998gg}. The function $\mu(\varphi)$ is unconstrained; it will be referred to as the \emph{mass aspect function} in what follows for reasons which will become clear below. The metric \eqref{banados} provides an exact solution of the time-symmetric constraint equations \emph{near} the conformal boundary.

 It is well known on which manifolds mass aspect functions with constant $\mu(\varphi) = m$ can be realised: $m>0$ gives time-symmetric slices of non-rotating BTZ black holes (also known as \emph{funnels} in hyperbolic geometry), $m=0$ yields the hyperbolic cusp, and $m<0$ corresponds to a manifold with one conical singularity. 
 
 The Hamiltonian mass of \eqref{banados}, measured with respect to the hyperbolic cusp, can be computed with a variety of methods \cite{Regge:1974zd, Chrusciel:2002cp, Kijowski:1979dj, Lee:1990nz, deHaro:2000vlm}, which all lead to 
 \begin{equation}
    H = \frac{1}{2 \pi} \int_{S^1} \mu(\varphi) d\varphi\,.
 \end{equation}
While in spatial dimension $n\geq 3$ either the Hamiltonian mass or the length of the energy-momentum vector are global invariants, this is not true anymore in $n=2$ dimensions \cite{Brown:1986nw}. This stems from the fact that the conformal group is finite-dimensional in $n\geq3$ dimensions, but infinite-dimensional in $n=2$ as it consists of arbitrary diffeomorphisms of the $S^1$.
Indeed, under an asymptotic symmetry transformation of the form\footnote{Here, we consider asymptotic symmetries leaving the initial data surface fixed.}
\begin{equation}
      (r, \varphi) \mapsto \left(
        \frac{r} {f^\prime(\varphi)},\, 
      f(\varphi) - \frac{f''(\varphi)}{2 r^2 } \right)
      \,.
  \end{equation}
where $f:S^1 \to S^1$ is any diffeomorphism of the circle, the
mass aspect function transforms as 
\begin{equation}
\label{trafofunction}
    \mu(\varphi) \mapsto \hat{\mu}(\varphi) = \mu(f(\varphi))f'(\varphi)^2 - 2 S(f)(\varphi)\,,
\end{equation}
where $S(f)$ denotes the Schwarzian derivative
  \begin{equation}
  {S(f)(\varphi)} =
    \frac{f^{(3)}( \varphi)}{f'(\varphi)} - \frac{3}{2} \left( \frac{f''(\varphi)}{f'(\varphi)}\right)^2
    \,,
    \label{10IX23.31}
  \end{equation}
and the Hamiltonian mass transforms as 
  \begin{equation}
       H \mapsto \hat{H} = \frac{1}{2\pi}
       \int_{S^1}
        \big(
        \mu(f({\varphi})) f'({\varphi})^2 - 2 S(f)(\varphi)
       \big)
        \,d\varphi\,.
  \end{equation}
 Upon varying $f$, $\hat{H}$ can be made arbitrarily large. Furthermore, if $\mu(\varphi) < -1$, there exists diffeomorphisms $f$ such that $\hat{H}$ is unbounded from below \cite{Balog, Barnich:2014zoa}. However, for $\mu(\varphi) \geq -1$, a global invariant can be obtained via minimisation
  \begin{equation}
        \underline{H}[\mu]:= \inf _{f}  H[\mu;f] \,,
        \label{minen}
      \end{equation}
 where for constant $\mu = m  \geq -1$, the infimum is attained on the identity map.
 One can go further than this and prove a positive mass theorem in two spatial dimensions \cite{Chrusciel:2024vle}. 
Indeed, in \cite{Chrusciel:2024vle} we applied the Witten spinorial method \cite{Witten:1981mf} to $n=2$, yielding either one or two inequalities for mass and angular momentum, depending on the topology of the manifold. Although the Witten spinorial method had already been applied to $n=2$  previously \cite{Cheng:2005wk, Chrusciel:2003qr}, our work addressed previously ignored  subtleties related to the definition of mass and inequivalent spin structures in two spatial dimensions.

 Interestingly, there exist mass aspect functions that cannot be mapped to a constant \cite{Balog} under an asymptotic symmetry. However, also in that case invariants may be obtained. Following \cite{Balog}, one may classify the function $\mu(\varphi)$ up to asymptotic symmetry transformations (where we now restrict ourselves to positively oriented diffeomorphism of $S^1$). It turns out that classifying such functions is equivalent to classifying $\mathrm{Diff}^+(S^1)$-inequivalent solutions to the \emph{Hill equation}. This can be see as follows: Consider the Hill equation
 \begin{equation}
    \frac{d^2 \psi}{d^2 \varphi} - \frac{\mu}{4} \psi = 0\,,
 \end{equation}
 where now $\varphi \in \mathbb{R}$, $\mu(\varphi)$ has been lifted to a  $2 \pi$-periodic function and $\psi : \mathbb{R} \rightarrow \mathbb{R}$. Upon changing $\varphi \mapsto \hat{\varphi} = f(\varphi)$, the function
   \begin{equation}
   \hat \psi (\varphi):= \frac{\psi(f\big(\varphi)\big)}{\sqrt{f'(\varphi)}}
  \end{equation}
  satisfies again a Hill equation, but now with $\hat \mu = 
               (f')^2 \,  \mu\circ f- 2 S(f)
               \,$, which is precisely of the form (\ref{trafofunction}).
    This can be exploited to classify the mass aspect functions $\mu$ up to transformations $\mu \mapsto  \hat{\mu}$ into different types.
    To determine the type, one needs the numbers of zeros of the solutions $\psi$ of the Hill equation, as well as the trace of the so-called \emph{Monodromy matrix}, defined as follows: Let $\Psi:= (\psi_1,\psi_2)$ be a basis of solutions. The periodicity of $\mu$ implies that 
     $\Psi(2\pi+\varphi)= \big(\psi_1(2\pi +\varphi),\psi_2(2\pi + \varphi) \big)$ is also a basis of solutions. 
   Hence there exists a matrix $\mono$, called monodromy matrix, such that
     $$
     \Psi (2\pi+\varphi) = \mono \Psi(\varphi)
     \,.
   $$
   Under a change of basis $\Psi \mapsto A \Psi$, the matrix $\mono$ changes as
   $
   \mono  \mapsto A \mono  A^{-1}
    \,,
   $
however, the trace of $\mono$ is an invariant. 
\section{Gluing}
A key direction of research in mathematical relativity is the construction of general relativistic data with interesting properties. In~\cite{ILS}, the so-called ``Maskit gluing'' technique was introduced, providing a way to generate new initial data sets with negative cosmological constant by performing a connected-sum construction at the conformal boundary. A different, localised version of this procedure was later developed in~\cite{ChDelayExotic}.
This gluing can be used to prove the hyperbolic positive energy theorem \cite{Chrusciel:2019cas}, as well as show the existence of certain asymptotically locally hyperbolic manifolds with constant scalar curvature and negative mass \cite{Chrusciel:2021ufc, Chrusciel:2022rlk}. 
A natural question is thus the behavior of mass under Maskit gluing.

In \cite{Chrusciel:2024pvn}, we analysed this question for vacuum and time-symmetric initial data sets in spatial dimension $n=2$. While there the gluing is simple, extracting the mass of the glued manifold is not, due to the complicated way the asymptotic symmetries act on the mass aspect function. 
The trick is to not only glue the ALH manifolds $(M_1, g_1)$, $(M_2, g_2)$, each of which has a well-defined mass aspect function $\mu_1(\varphi)$ and $\mu_2(\varphi)$, respectively, but also to glue their associated basis of solutions to the Hill equation $\Psi_1$, $\Psi_2$. In this way one gets a glued ALH manifold $(M, g)$ together with an associated glued basis of solutions to the Hills equation $\Psi$. From this, one can determine the mass aspect function $\mu(\varphi)$ of the glued manifold by employing the classification \cite{Balog}.

This leads to the following theorem \cite{Chrusciel:2024pvn}:
  \begin{Theorem}
    Given two asymptotically locally hyperbolic manifolds in dimension $n = 2$ with constant scalar curvature and positive initial masses $m_1$ and $m_2$, the glued manifold has mass $m$ determined from the equation
\begin{equation}
  \hspace{-0.9cm}
     \cosh(\sqrt{m}\pi) 
    =   2 \omega_1
  \omega_2 \cosh
  (\sqrt{m_1} \pi) \cosh
  (\sqrt{m_2} \pi )- \cosh
  (\sqrt{m_1} \pi-\sqrt{m_2} \pi )
  \nonumber
\end{equation}
with gluing parameters $\omega_1>1$, $\omega_2>1$.
\end{Theorem}
 Similar theorems can be found in \cite{Chrusciel:2024pvn} for gluing any possible combination of asymptotically locally hyperbolic manifolds with constant mass aspect functions. This can be used to show
\cite{Chrusciel:2024pvn}:
  \begin{Theorem}
   All mass aspect functions can be realised by smooth asymptotically locally  
       hyperbolic constant scalar curvature manifolds, which have
        at most one conical singularity.
\end{Theorem}
The theorem in particular demonstrates on which manifolds mass aspect functions which \emph{cannot} be mapped to a constant can be realised.
\section*{Acknowledgements}
I acknowledge support by the Heising-Simons Foundation under the ``Observational Signatures of Quantum Gravity'' collaboration grant 2021-2818 and the U.S. Department of Energy, Office of High Energy Physics,
under Award No. DE-SC0019470.

\bibliography{ChruscielWutte-minimal}

\providecommand{\newblock}{}
\begin{thebibliography}{10}
\expandafter\ifx\csname url\endcsname\relax
  \def\url#1{{\tt #1}}\fi
\expandafter\ifx\csname urlprefix\endcsname\relax\def\urlprefix{URL }\fi
\providecommand{\eprint}[2][]{\url{#2}}

\bibitem{Maldacena:1997re}
Maldacena J~M 1998 {\em Adv. Theor. Math. Phys.\/} {\bf 2} 231--252 (\textit{Preprint} \eprint{hep-th/9711200})

\bibitem{BartnikIsenberg}
Bartnik R and Isenberg J 2002 {The Constraint equations} {\em {50 Years of the Cauchy Problem in General Relativity: Summer School on Mathematical Relativity and Global Properties of Solutions of Einstein's Equations}\/} (\textit{Preprint} \eprint{gr-qc/0405092})

\bibitem{CarlottoLRR}
Carlotto A 2021 {\em Living Rev. Rel.\/} {\bf 24} 2

\bibitem{Banados:1992wn}
Ba{\~n}ados M, Teitelboim C and Zanelli J 1992 {\em Phys. Rev. Lett.\/} {\bf 69} 1849--1851 (\textit{Preprint} \eprint{hep-th/9204099})

\bibitem{Brown:1986nw}
Brown J~D and Henneaux M 1986 {\em Commun.\ Math.\ Phys.\/} {\bf 104} 207--226

\bibitem{Chrusciel:2024pvn}
Chru{\'s}ciel P~T and Wutte R 2024  (\textit{Preprint} \eprint{2401.04048})

\bibitem{Chrusciel:2024vle}
Chru{\'s}ciel P~T, Cong W, Qu{\'e}au T and Wutte R 2025 {\em Class. Quant. Grav.\/} {\bf 42} 085010 (\textit{Preprint} \eprint{2411.07423})

\bibitem{Skenderis:1999nb}
Skenderis K and Solodukhin S~N 2000 {\em Phys. Lett. B\/} {\bf 472} 316--322 arXiv:hep-th/9910023 (\textit{Preprint} \eprint{hep-th/9910023})

\bibitem{Banados:1998gg}
Ba{\~n}ados M 1999 {\em AIP Conf. Proc.\/} {\bf 484} 147--169 arXiv:hep-th/9901148 (\textit{Preprint} \eprint{hep-th/9901148})

\bibitem{Regge:1974zd}
Regge T and Teitelboim C 1974 {\em Annals Phys.\/} {\bf 88} 286

\bibitem{Chrusciel:2002cp}
Chru\'sciel P~T, Jezierski J and Kijowski J 2002 {\em {Hamiltonian field theory in the radiating regime}\/} vol~70

\bibitem{Kijowski:1979dj}
Kijowski J and Tulczyjew W~M 1979 {\em A symplectic framework for field theories\/} Lecture notes in physics; 107 (Berlin: Springer-Verlag) ISBN 0387095381

\bibitem{Lee:1990nz}
Lee J and Wald R~M 1990 {\em J. Math. Phys.\/} {\bf 31} 725--743

\bibitem{deHaro:2000vlm}
de~Haro S, Solodukhin S~N and Skenderis K 2001 {\em Commun. Math. Phys.\/} {\bf 217} 595--622 (\textit{Preprint} \eprint{hep-th/0002230})

\bibitem{Balog}
Balog J, Feher L and Palla L 1998 {\em Int.\ Jour.\ Mod.\ Phys.\ A\/} {\bf 13} 315--362 arXiv:hep-th/9703045 (\textit{Preprint} \eprint{hep-th/9703045})

\bibitem{Barnich:2014zoa}
Barnich G and Oblak B 2014 {\em Class.\ Quantum Grav.\/} {\bf 31} 152001 arXiv:1403.3835 [hep-th] (\textit{Preprint} \eprint{1403.3835})

\bibitem{Witten:1981mf}
Witten E 1981 {\em Commun. Math. Phys.\/} {\bf 80} 381

\bibitem{Cheng:2005wk}
Cheng M~C~N and Skenderis K 2005 {\em JHEP\/} {\bf 08} 107 (\textit{Preprint} \eprint{hep-th/0506123})

\bibitem{Chrusciel:2003qr}
Chru\'sciel P~T and Herzlich M 2003 {\em Pacific J. Math.\/} {\bf 212} 231--264

\bibitem{ILS}
Isenberg J, Lee J~M and Stavrov~Allen I 2010 {\em Ann.\ Henri Poincar\'e\/} {\bf 11} 881--927 ISSN 1424-0637 arXiv:0910.1875 [math-dg] \urlprefix\url{http://dx.doi.org/10.1007/s00023-010-0049-0}

\bibitem{ChDelayExotic}
Chru{\'s}ciel P~T and Delay E 2018 {\em J. Diff. Geom.\/} {\bf 108} 243--293 (\textit{Preprint} \eprint{1511.07858})

\bibitem{Chrusciel:2019cas}
Chru{\'s}ciel P~T and Delay E 2019  (\textit{Preprint} \eprint{1901.05263})

\bibitem{Chrusciel:2021ufc}
Chru{\'s}ciel P~T, Delay E and Wutte R 2023 {\em Adv. Theor. Math. Phys.\/} {\bf 27} 1333--1403 (\textit{Preprint} \eprint{2112.00095})

\bibitem{Chrusciel:2022rlk}
Chru{\'s}ciel P~T and Delay E 2023 {\em SIGMA\/} {\bf 19} 005 (\textit{Preprint} \eprint{2207.14563})

\end{thebibliography}
\end{document}